# Nanogap-enhanced Infrared Spectroscopy with Template-stripped Wafer-scale Arrays of Buried Plasmonic Cavities

*Xiaoshu Chen,[1] Cristian Ciracì,[2,3] David R. Smith,[3] and Sang-Hyun Oh*,[1]*

[1]Department of Electrical and Computer Engineering, University of Minnesota,

Minneapolis, Minnesota 55455 USA.

[2]Istituto Italiano di Tecnologia (IIT), Center for Biomolecular Nanotechnologies,

Via Barsanti, I-73010 Arnesano, Italy.

[3]Center for Metamaterials and Integrated Plasmonics,

Department of Electrical and Computer Engineering,

Duke University, Durham, North Carolina 27708, USA.

Corresponding author *E-mail: sang@umn.edu

**ABSTRACT**

We combine atomic layer lithography and template stripping to produce a new class of substrates for surface-enhanced infrared absorption (SEIRA) spectroscopy. Our structure consists of a buried and U-shaped metal-insulator-metal waveguide whose folded vertical arms efficiently couple normally incident light. The insulator is formed by atomic layer deposition (ALD) of $Al_2O_3$ and precisely defines the gap size. The buried nanocavities are protected from contamination by a silicon template until ready for use and exposed by template stripping on demand. The exposed nanocavity generates strong infrared resonances, tightly confines infrared radiation into a gap that is as small as 3 nm ($\lambda/3300$), and creates a dense array of millimeter-long hotspots. After partially removing insulators, the gaps are backfilled with benzenethiol molecules, generating distinct Fano resonances due to a strong coupling with gap plasmons, and SEIRA enhancement factor of $10^5$ is observed for a 3-nm gap. Because of wafer-scale manufacturability, single-digit-nanometer control of the gap size via ALD, and long-term storage enabled by template stripping, our buried plasmonic nanocavity substrates will benefit broad applications in sensing and spectroscopy.

**KEYWORDS.** Surface-enhanced infrared absorption (SEIRA), nanogap, template stripping, surface plasmon, gap plasmon, atomic layer deposition, Fano resonance, nanocavity

Nanometer-scale gaps in noble metals are one of the key building blocks for plasmonic devices.[1-4] Plasmonic nanostructures that are characterized by small gaps can concentrate electromagnetic waves into nanometric volumes, enhancing light-matter interactions in an unprecedented way.[5,6] This unique capability is utilized for fundamental investigation of nanophotonic phenomena, exposing light-induced electron tunneling[7,8] and nonlocality,[9] as well as applications in surface-enhanced spectroscopies[1,3,10,11] optical trapping,[12,13] and nonlinear optics.[14] Point-like nanogaps formed by nearly touching metal nanoparticles have been extensively studied in the context of surface-enhanced Raman spectroscopy (SERS).[7] By using flat-sided nanoparticles such as nanocubes or thin-film stacks, it is possible to create extended gaps forming metal-insulator-metal (MIM) waveguides[2] that can support highly confined electromagnetic modes—gap plasmons[15]—that propagate with no cut-off. As the gap size is reduced to single-digit nanometers, and the particle dimensions are kept constant, the resonances of such structures shift toward longer wavelengths, typically out into the near-infrared. Strong plasmon resonances can in principle also be obtained in the mid-infrared (mid-IR) regime, where lower optical losses than in the visible and near-infrared, make them well-suited for applications such as thermal emission[16] and surface-enhanced infrared absorption (SEIRA).[17,18]

Infrared (IR) absorption spectroscopy is an important analytical technique that complements Raman spectroscopy to acquire molecular vibrational fingerprints in a label-free manner. The small IR absorption cross-section of molecules, however, presents a challenge for sensitive detection. In SEIRA, surface-plasmon-enhanced local optical fields boost IR absorption of molecules adsorbed on metallic surfaces. Various geometries such as metallic islands,[17] nanorods,[19-22] nanoholes,[23] split rings,[24] and nanoshells,[7] are utilized to enhance local electromagnetic fields. While isolated metallic structures have shown impressive performance for SEIRA, coupled nanogap structures that take advantage of gap plasmons possess an ideal geometry to maximize field confinement and enhancement – as has been shown for SERS. Plasmonic nanogap structures have been fabricated using electron-beam lithography,[25-27] focused ion beam (FIB),[28,29] electromigration,[3] and nanosphere lithography.[30-32] Reproducible manufacturing of single-digit nanometer gaps, however, remains a challenge. For broader dissemination, improved sensitivity, reproducibility, and to relax the requirement of intense sources for SEIRA, it is important to manufacture large-area substrates with dense hotspot arrays.

In addition to direct-write lithography, nanogaps can also be constructed via thin-film deposition.[2,30,33] The key advantage of this approach is that the film thickness can readily be controlled with single-digit-nanometer- or even Ångstrom-scale resolution using atomic layer deposition (ALD). A new patterning approach based on ALD – atomic layer lithography – can create gaps with sub-nm thickness.[34] The present work combines atomic layer lithography with template stripping, which uses silicon templates to produce smooth patterned metals,[35-37] to construct strippable plasmonic nanocavities for SEIRA and achieve strong field enhancements through precise control over the gap size.

As illustrated in Figure 1, our MIM nanocavity is constructed by burying dielectric-coated metal patterns (e.g. stripes, disks) into another metal film. As a result, the nanocavity is folded into a U-shape (Fig. 1) wherein two vertical arms efficiently couple the incident light. The resultant buried cavity geometry is mechanically robust against wet processing and present planarized top surfaces that facilitate subsequent integration with other devices or microfluidics. The insulating film in the vertical arms can be removed without compromising the mechanical stability of the cavity and backfilled with analyte molecules for enhanced sensing. Plasmonic hotspots are generated along the entire length of a 1.5-mm long nanocavity, thus our structure is robust against local defects or impurities when compared to point-like nanogap structures. Importantly, these cavities are protected by a silicon template, and can be exposed via template stripping immediately before use, preventing surface contamination. We show SEIRA spectroscopy results using these structures and distinct Fano resonances due to a strong coupling between molecules and gap plasmons.

The process flow for making our nanogap structure is illustrated in Figures 1a-e. First, 80-nm-thick gold stripes are patterned on a 4-inch Si wafer using standard photolithography followed by metal evaporation and lift-off. Note that no adhesion layer is used between the gold film and Si substrate, a necessity for template-stripping of the final structure. The gold stripes are then conformally coated with a thin layer of alumina ($Al_2O_3$) using ALD (Fig. 1b). A second layer of metal (150-nm-thick silver) is deposited on top of the alumina forming the nanogap between the gold and silver films, with the gap width precisely defined by the thickness of ALD-grown $Al_2O_3$ film (Fig. 1c). Finally, the nanogap structures are template-stripped and exposed for molecule deposition and spectroscopic measurements. For this template-stripping step, a UV-curable optical adhesive (NOA 61, Norland Products Inc.) is applied to the surface of a silver film (Figure 1d), covered with a glass slide, cured under a UV lamp and on a hotplate, and the whole structure is stripped off of the Si wafer (Fig. 1e). A schematic cross section of the buried nanocavity device is illustrated in Figure 1f. We imaged the sample surface of a 5-nm gap pattern with atomic force microscopy. A line scan (Fig. 1g) shows the height difference, due to the thickness of the alumina layer, across template-stripped gold and silver surfaces. This height difference arises from the relatively stronger adhesion of alumina to silicon than to silver, the alumina film on the flat region is not peeled off during the template-stripping process. Only the alumina film inside the gap is transferred with the sample.

Using standard photolithography, any pattern shapes can be produced over a large area (Fig. 1h). The scanning electron micrograph (SEM) in Fig. 1i, which was taken after template stripping, shows cavities with a 5-nm gap size. Because the nanogaps are made entirely with wafer-scale batch processes, namely metal deposition, photolithography and ALD, there is no limitation in the size and shape of the cavity patterns. It is straightforward to produce a densely packed array of millimeter- or centimeter-long nanocavities over an entire 4-inch Si wafer in any shape, such as disks (Fig. 1j) or wedges (Fig. 1k). The high packing density combined with millimeter-scale horizontal length (perpendicular to the cavity length direction) facilitates FTIR measurements. A Nicolet Series II Magna-IR System 750 FTIR equipped with an IR microscope (15× IR objective lens, NA=0.58) was used to measure the spectra from the nanogap samples in reflection mode using unpolarized light. The spectrometer and the microscope were purged with dry air. The sample and the light path from the objective lens to the sample are exposed in atmosphere. An adjustable built-in aperture in the IR microscope is used to define the illumination spot size. We measured absorbance spectra (defined as $\log_{10}(I_o/I)$, where $I_o$ is reflected signal from a reference bare silver surface and I is reflected signal from nanogap cavities.) from nanogap cavities with metal stripe width of 500, 600, 700, 900, 1300, and 1700 nm, each with gap sizes of 3, 5, 7, and 10 nm. The gold film thickness is fixed at 80 nm for all of the samples. Figure 2a shows the spectra measured from 3-nm gap structures, with different cavity lengths. Multiple Fabry-Pérot (FP) resonances are observed at wavelengths between 10 $\mu$m and 1.6 $\mu$m. Similar FP resonances are also observed in disk-shaped nanocavities (data not shown).

The peaks observed in the spectra shown in Fig. 2a correspond to the FP resonances in the buried nanogap cavity.[2,38] While the cavity is folded, the gap plasmon follows the dispersion as if it was propagating in an unfolded nanogap cavity.[39] Although we use unpolarized light to launch gap plasmon modes in our structures, only the TM component (electric field perpendicular to the stripes) couples to the gap-plasmon mode in the 2D cavities. Moreover, with normally incident light, only symmetric FP modes (with respect to the magnetic field) can be excited. In our experiments, anti-symmetric modes are also observed as small peaks between much stronger resonance peaks associated with symmetric modes, because the incident light is not perfectly collimated (NA=0.58). Simulation results show that the leftmost peak of each spectrum in Figure 2a is the first symmetric FP mode.

As with other MIM cavity structures, spectral scaling can be simply achieved by tuning cavity length or the gap thickness. Increasing the cavity length shifts the FP modes toward longer wavelengths. With micron-scale cavity lengths fabricated using photolithography, the cavity resonance can be tuned from the mid- to near-infrared. Sub-micron-scale cavities fabricated via electron-beam lithography can possess resonances in the visible regime, which will be the focus of future work. Many high-order FP modes are observed from longer cavities. Figure 2b shows a reflection spectrum obtained from a 5-nm gap cavity, plotted as a function of wavelength, which shows a series of sharp resonances peaks. A Q factor – defined as $\lambda_{resonance}/\Delta\lambda$ – of 34 was measured at a resonance wavelength of 1.8 $\mu$m. For a decreasing gap thickness, these resonances shift toward longer wavelengths because of the gap-plasmon dispersion.[40] That is, as the gap shrinks, the effective refractive index of the tightly confined plasmonic mode grows, increasing the effective cavity length and thus the resonant wavelength.

Using the peak resonance wavelength, we can deduce the dispersion characteristics of the gap plasmon in buried nanocavities, as shown in Figure 2c for the gap sizes of 3, 5, 7, and 10 nm. The results from an analytical dispersion equation[2,41] of 2D gap plasmons are shown as the solid lines in Fig. 2c. For the wider gaps (7 and 10 nm in width), both theory and experimental data are in good agreement. For 3 and 5 nm gaps, the experimental data deviate toward shorter wavelengths compared with theoretical predictions, likely due to the roughness of the metal surface in the nanogap. For the 3-nm gap, we find an effective refractive index of 6.5 for the gap plasmon mode. The resonance peak measured at 10.16 $\mu$m wavelength from this sample indicates that free-space radiation can be squeezed into a cavity as narrow as 3 nm ($\lambda$/3300), demonstrating the strong optical confinement enabled by the MIM geometry.

It is interesting to note that the spectra shown in Figure 2a show very distinct peak resonances, which is indicative of good coupling between the gap plasmon and far-field modes. This coupling is quite remarkable for the structure consisting of bare metal for more than 99.5% of its topographically flat surface. The strong coupling can be understood intuitively by replacing the fields radiated by the slits (formed by the vertical arms of the cavity) with effective magnetic currents that re-emit a wave whose field has the opposite phase to the scattered wave reflecting off the metal region. The resultant destructive interference reduces the total reflected fields, greatly enhancing the energy absorbed from the incident wave.[40,42]

After mapping the resonances of the nanocavity structures, we coated the exposed cavities with a monolayer of benzenethiol (BZT) molecules to perform nanogap-enhanced IR absorption spectroscopy. Finite-element method (FEM) calculations using COMSOL, a commercially available finite-element based solver, predict that the strongest electric field of the first FP mode exists at the end facets of the nanogap cavity, with a simulated field intensity ($|E|^2$) enhancement of about 1600 (Figure 3a). To utilize the maximum near-field strength inside the gap, molecules should be placed inside the gap region, which is initially filled with the $Al_2O_3$ film. To accomplish this, Buffered Oxide Etch (BOE) is used to partially remove $Al_2O_3$ inside the nanogap (see Supporting Information for details). After cleaning with DI water and drying with nitrogen, the sample is soaked in a 2 mM benzenethiol solution in ethanol for 24 hrs to coat the exposed gold and silver surfaces with a monolayer of BZT.[43] Excess BZT is removed by cleaning with ethanol. The spectra from BZT-coated samples are then measured by FTIR (Figure 3b). Six absorption peaks in the mid-IR for BZT – 1000, 1022, 1073, 1181, 1473, and 1575 cm$^{-1}$ – are in the resonant range of our nanogap structures, which falls between 10 $\mu$m to 6 $\mu$m. The first-order resonance of our device can be tuned from 6 $\mu$m to 11.5 $\mu$m (with a measured spectral linewidth less than 1 $\mu$m) by changing the width of the metal stripes from 450 nm to 900 nm. Enhancement factors (EF) of up to $10^5$, which is comparable to EFs measured from other high-performance SEIRA substrates, are calculated using experimental data (details in Supporting Information). These EFs are also in the range predicted by our numerical simulations if the normalization of the field enhancement is calculated with respect to the local fields at a silver surface (usually < 10%)."

The measured spectra show an asymmetric Fano-shape resonance[19,21,44,45] for each absorption band, a behavior confirmed by numerical calculations. The BZT monolayer can be modeled as a thin layer whose dielectric function is obtained as a sum of Lorentz oscillators that match the absorption resonances of the BZT molecule. For the sake of clarity, however, we considered just one oscillator corresponding to the main resonance at 1473 cm$^{-1}$. As for the actual experimental sample, we intercalated BZT roughly 3 nm into the gap (controlled by the BOE processing time) to increase the coupling with the gap-plasmon mode of the structure. A set of reflection spectra calculated for different stripe widths is shown in Figure 3c. As the resonance of the system is shifted across the BZT absorption resonance, very distinct Fano-shape features appear in the reflection spectra. These resonances are qualitatively identical to those shown in Figure 3b depending on the relative position of the BZT absorption peak with respect to the gap-plasmon resonance.

By combining nanogaps with different cavity lengths, it is possible to create a broadband plasmonic resonator that covers a wide spectral region of the mid-IR fingerprint region. Similar ideas have been used by others to obtain broadband resonances from engineered optical antennas.[46,47] As shown in Figure 4a, we made an array of metal stripes with the widths ranging from 500 nm to 1010 nm in 30 nm steps). The spectrum from this laterally graded cavity resonator is shown in Figure 4b. Resonances from nanogap cavities consisting of uniform metal stripe widths of 500, 600, 700, 900 and 1300 nm are also plotted in the same figure. The device with laterally graded nanocavities possesses a resonance 5-6 times wider than the single cavity length resonators. We confirmed this observation by simulating a nanogap cavity with metal stripe widths from 500 to 1000 nm with 50 nm steps, as shown in Figure 4c. We found that the first resonance mode from each device spans a wavelength range from 11 $\mu$m to 6 $\mu$m, covering the typical mid-IR fingerprinting region for molecules. A laterally graded nanogap structure with a 3 nm gap size was used for BZT sensing to demonstrate the functionality of the broadband resonator（Figure 4d）.

In summary, we combined atomic layer lithography and template stripping to fabricate buried plasmonic nanocavity arrays at the wafer scale with a minimum gap size of 3 nm. These new class of substrates for SEIRA can generate intense gap plasmons that boost coupling of IR radiation with molecules situated in the gaps. High SEIRA enhancements of up to 5 orders of magnitude are observed. As the hotspots of each nanocavity uniformly extend along millimeter-long lines, our substrates are robust against localized defects or contaminants, which will enable more reproducible SEIRA spectroscopy and also combine SERS and SEIRA on the same substrate.[48] As with other template-stripped metallic nanostructures,[49] the surfaces and high-field area of these strippable nanocavities are initially protected by a silicon template, which provides a robust protection mechanism against contamination or degradation, and thus each chip can be stored for an extended time and template-stripped on-demand. Smaller gap dimensions (down to 1 nm) are achievable using atomic layer lithography[34] even though they were not needed for the molecular sensing experiments performed here. Because the Au and Ag films are electrically isolated by an insulating layer, applying a bias across the gap can turn these structures into vertically oriented tunnel junctions. Such structures can be used for studying electron tunneling,[50] nonlocal electrodynamics,[9] as well as for photodetection.[51]

**Supporting Information**

Detailed fabrication methods, FTIR instrument and measurement, surface functionalization, and calculation of SEIRA enhancement factors. This material is available free of charge via the Internet at http://pubs.acs.org.

Notes. The authors declare no competing financial interest.

**ACKNOWLEDGEMENT**

This research was supported by the U.S. Department of Defense (DARPA Young Faculty Award, #N66001-11-1-4152; X.S.C. and S.-H.O.), Office of Naval Research Young Investigator Program (S.-H.O.), and the Air Force Office of Scientific Research (AFOSR, Grant No. FA9550-12-1-0491; C.C. and D.R.S.) and the Army Research Office through a Multidisciplinary University Research Initiative (Grant No. W911NF-09-1-0539; C.C. and D.R.S.). The team also utilized resources at the University of Minnesota, including the Nanofabrication Center, which receives partial support from NSF through the National Nanotechnology Infrastructure Network, and the Characterization Facility, which has received capital equipment from NSF MRSEC. X.S.C. acknowledges support from the 3M Science and Technology Fellowship and the University of Minnesota Doctoral Dissertation Fellowship. The authors thank Jonah Shaver for helpful comments.

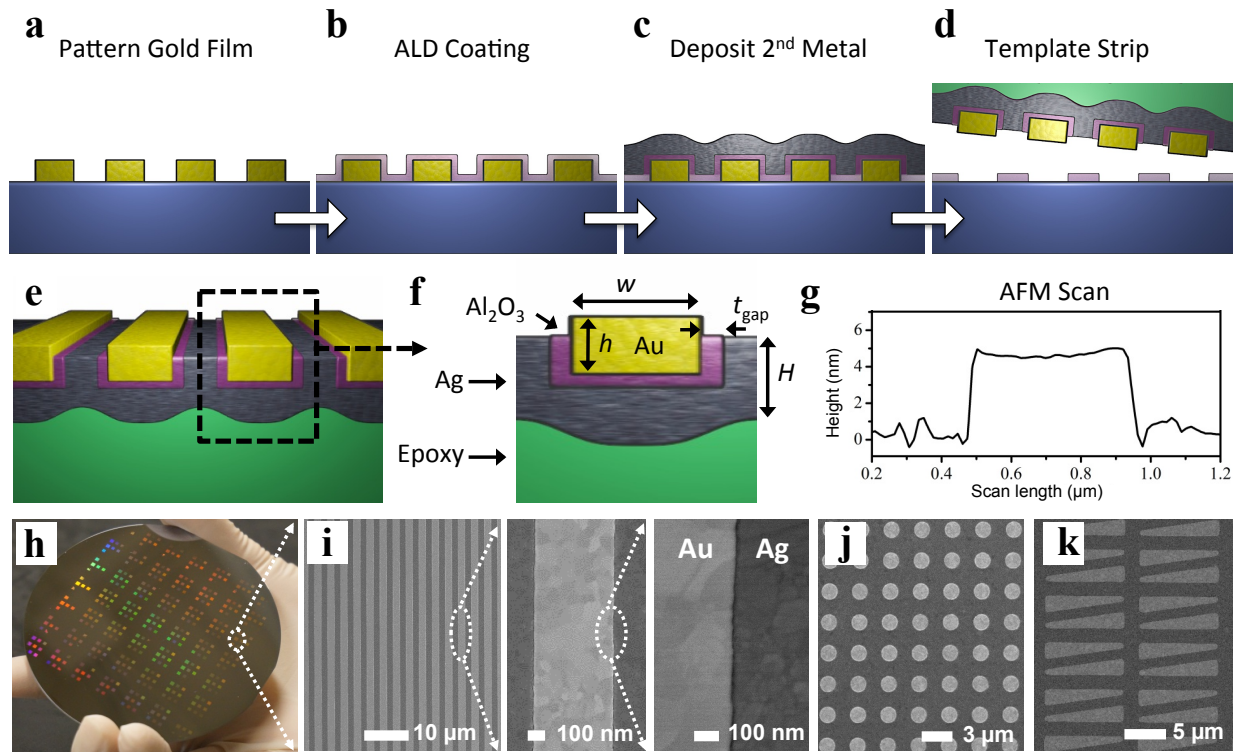

**Figure 1. a-e**, Schematic of the device fabrication process. Standard photolithography is used to pattern gold films on a 4-inch silicon wafer. These patterns are conformally encapsulated with a thin alumina spacer using atomic layer deposition (ALD). Next, a silver film is deposited conformally on the pattern, and the whole structure is stripped off from the silicon substrate using UV cured epoxy and a glass slide. **f,** Cross-sectional schematic of a buried nanocavity. **g,** Contact mode AFM line scan across a 5-nm nanogap cavity showing a height difference between the gold and silver films due to the 5-nm-thick $Al_2O_3$ film. **h,** Photograph of a 4-inch wafer containing metal stripes after lift-off. Each square is approximately 1.5 mm by 1.5 mm. **i**. SEM image of an array of buried nanogaps on a chip. Further zoomed-in images show a single cavity and a 5 nm nanogap on one side of the cavity. **j** and **k,** SEM of buried disks and wedges.

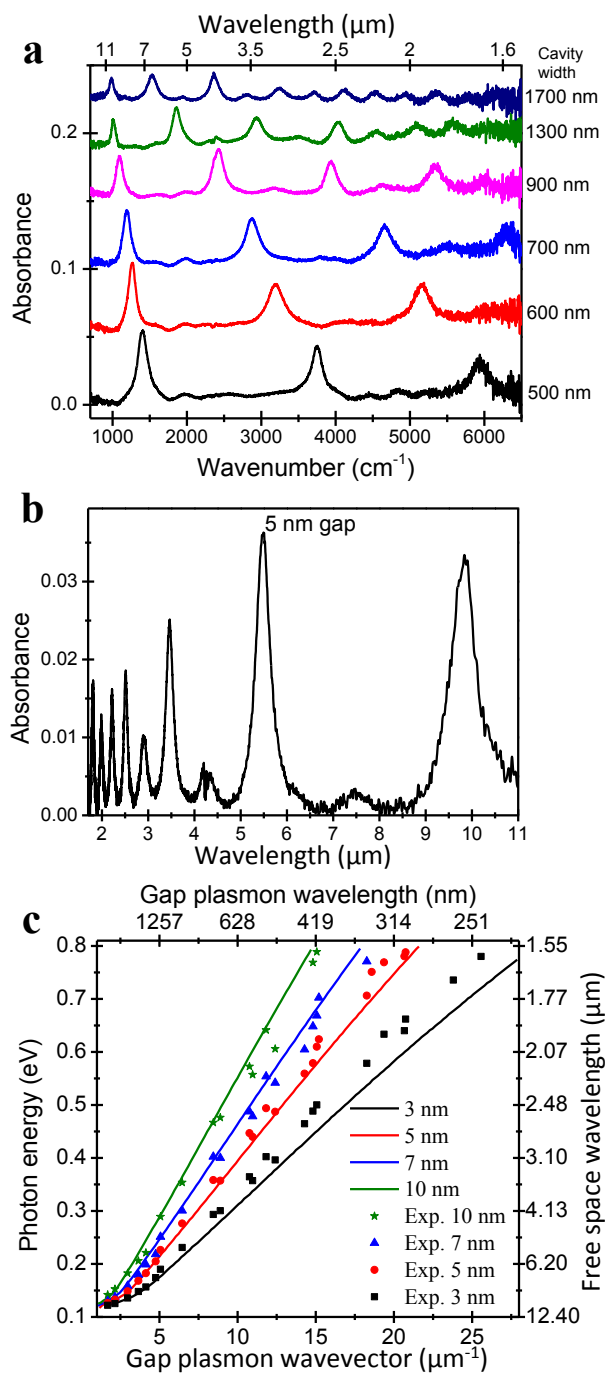

**Figure 2.** Mid-infrared spectra and dispersion of gap plasmons. **a**, FTIR spectra from 3-nm-gap cavities in a 2D stripe pattern with different metal stripe widths. Spectra are offset in y-axis for clarify. See the SI for measured spectra from 5, 7, and 10 nm nanogap cavities. **b**, FTIR spectrum measured from a 5-nm-gap cavity in a 2D stripe pattern with cavity width of 1700 nm, plotted in wavelength. **c**, Dispersion of nanogap from experimental and simulation data for 3, 5, 7, and 10

nm gaps. The solid lines are plotted using the theoretical dispersion curve[2] (see Supporting Information) of gap plasmons for different cavity lengths and gap sizes.

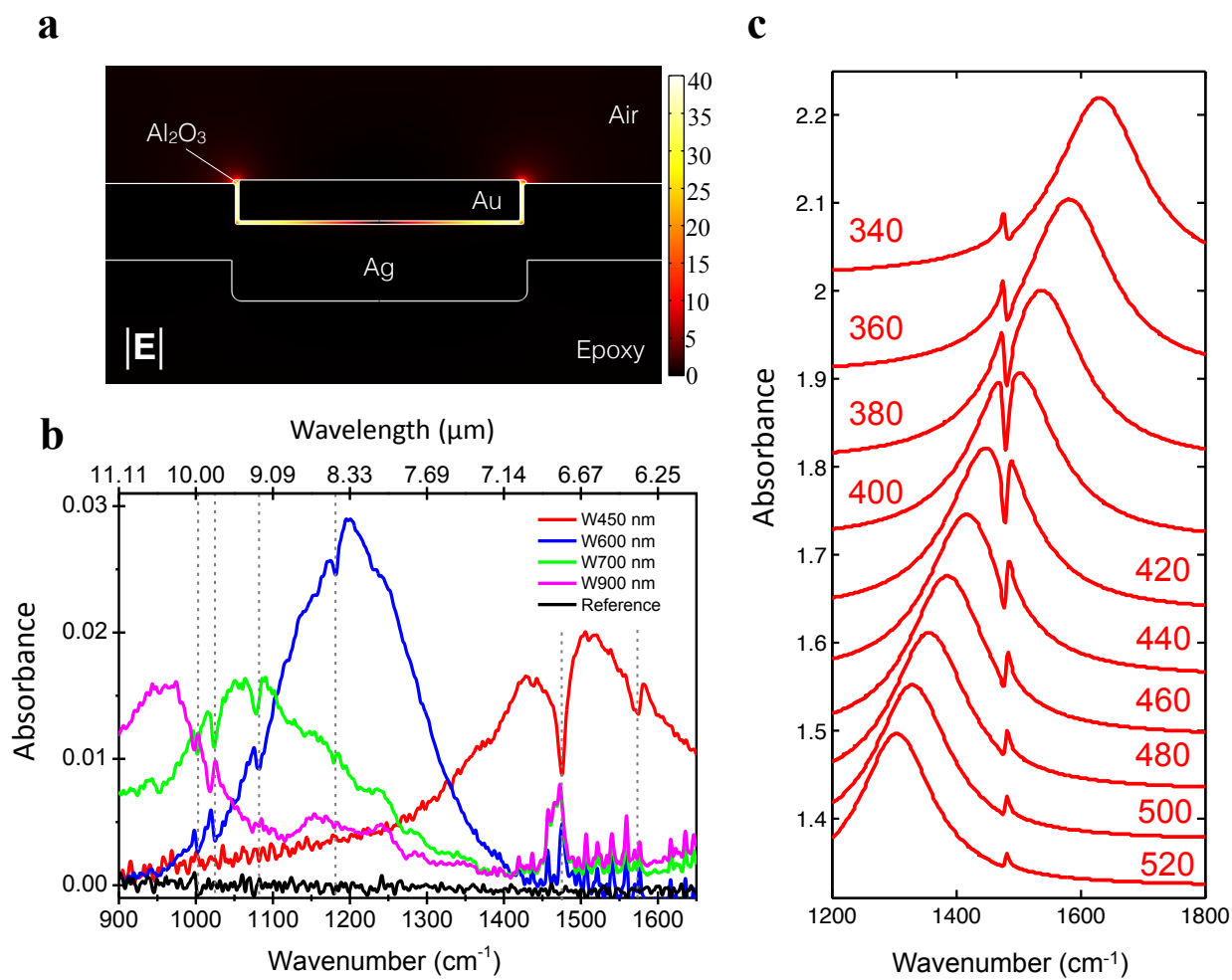

**Figure 3. a**, Field distribution in the buried nanogap. The field strength is highest at the nanogap outlets. **b**, Sensing of benzenethiol using a buried, 3 nm gap with various stripe widths. The black curve is the absorption of BZT monolayer coated on template stripped (TS) silver film without any pattern. Note that the absorption from BZT monolayer is below noise level. The vertical gray dashed lines in the figure indicate the absorption bands of BZT. **c**, Simulated Fano coupling between BZT absorption and gap plasmon resonance in nanogap devices with various metal stripe widths ranging from 340 nm to 520 nm.

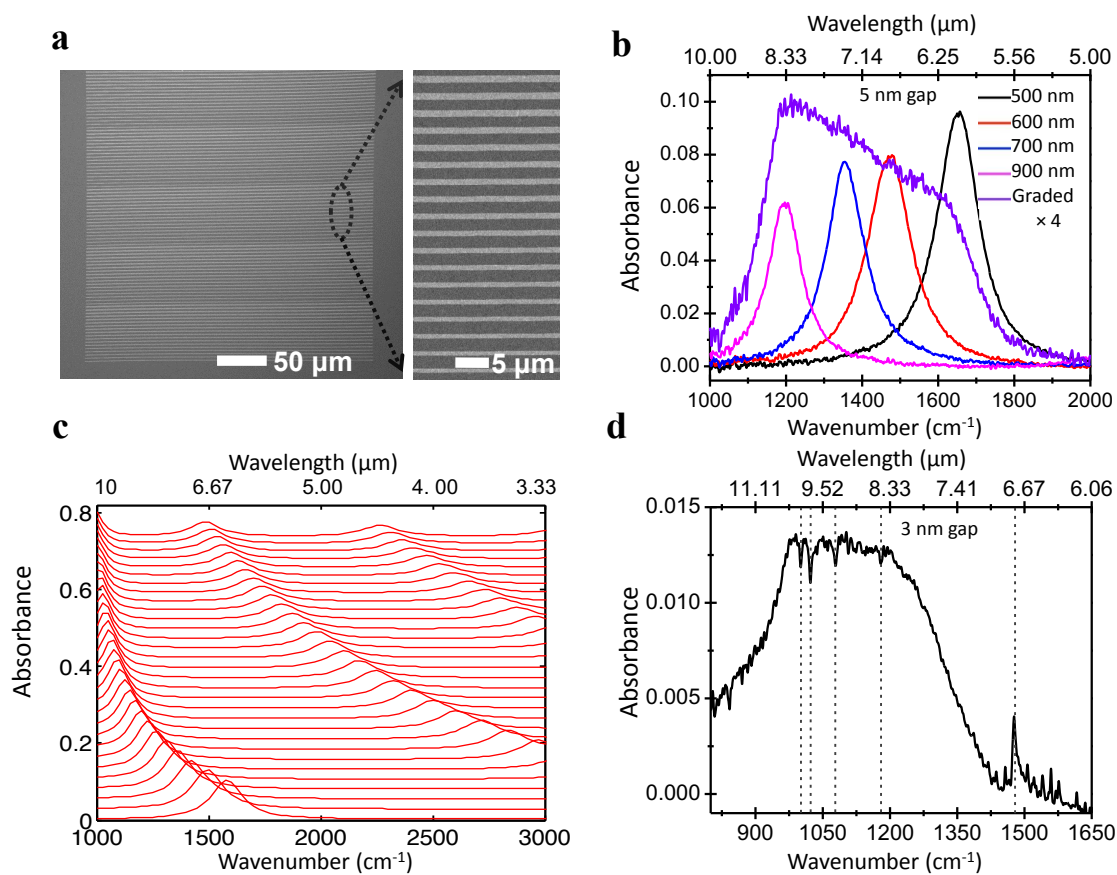

**Figure 4. Tunable broadband resonators. a**, SEM of a nanocavity array with varying metal stripe widths. Zoomed-in image shows a section of the device wherein the metal stripe (brighter region) width increases from 500 nm (bottom) to 1010 nm (top). **b**, FTIR spectra (purple line) from nanogap cavities with metal stripe width ranging from 500 nm to 1010 nm in 30 nm steps on a single device (multiplied by 4 for comparison). Spectra from 5 nm gaps with single width (500, 600, 700, and 900 nm) on one device (black, red, blue, and pink line, respectively) are plotted in the same figure. The spectra were measured from an area of approximately 130 $\mu$m by 130 $\mu$m. **c**, Simulation of a 5 nm nanocavity with various metal stripe widths. **d**, Sensing of benzenethiol using the a graded nanogap structure with a 3 nm gap size. Grey dashed lines indicate BZT absorption peaks.

# Supporting Information

# Nanogap-enhanced Infrared Spectroscopy with Template-stripped Wafer-scale Arrays of Buried Plasmonic Cavities

*Xiaoshu Chen,[1] Cristian Ciracì,[2,3] David R. Smith,[3] and Sang-Hyun Oh*[*,1]*

[1]Department of Electrical and Computer Engineering, University of Minnesota,

Minneapolis, Minnesota 55455 USA.

[2]Istituto Italiano di Tecnologia (IIT), Center for Biomolecular Nanotechnologies,

Via Barsanti, I-73010 Arnesano, Italy.

[3]Center for Metamaterials and Integrated Plasmonics,

Department of Electrical and Computer Engineering,

Duke University, Durham, North Carolina 27708, USA.

*To whom correspondence should be addressed: sang@umn.edu

**Detailed nanogap array fabrication method.**

Patterns of trenches with various sizes and shapes were made using a Canon 2500 i3 stepper and positive resist (SPR 955 – CM 0.7) over a whole 4-inch silicon wafer. Then 80 nm of gold was deposited on the pattern with electron beam evaporation (Temescal) without an adhesion layer. The wafer was then soaked upside down in acetone for 2 hours for lift-off, cleaned with methanol and isopropyl alcohol, and dried with nitrogen gun leaving gold stripes or disks on the silicon wafer. The wafer was then cut into pieces for alumina coating with different thicknesses, depending on application, using atomic layer deposition (Savannah, Cambridge Nano Tech Inc.) at a typical deposition rate 1~2 Å per cycle at 250 ºC. Immediately after alumina coating, the samples were transferred to an electron beam evaporator with a planetary fixture (CHA SEC 600) for conformal metal deposition (150-nm-thick silver). After this second metal deposition, UV-cure optical adhesive (NOA 61, Norland Products Inc.) was applied to the sample surface, which was in turn covered by a glass slide. After curing under UV light for 15 minutes and annealing on a hotplate at 55°C overnight, the whole pattern including both metal layers and the alumina layer in between was stripped off from the silicon wafer, exposing vertically oriented nanogaps.

**FTIR instrument and measurement.**

The instrument used for infrared absorption measurement was a Nicolet Series II Magna-IR System 750 FTIR equipped with a reflection mode microscope. All the data from the nanogap cavities were taken using the microscope with a liquid-nitrogen-cooled MCT-A (Mercury Zinc Telluride Alloy) detector. The infrared aperture size was around 130 μm by 130 μm. Each spectrum from the nanogap cavity was normalized to the signal taken from bare silver hundred of microns away from the patterned area using the same aperture size, and was averaged for 256 times with a resolution of 4 cm$^{-1}$ in each spectrum. To plot the theoretical plasmon dispersion curve, we used the gap plasmon dispersion equation:

$$|2LK_{MIM}(\omega) + \Delta\Phi_1 + \Delta\Phi_2| = 2\pi m,$$

where $K_{MIM}$ is the gap plasmon wavevector, $\Delta\Phi_1$ and $\Delta\Phi_2$ are the phase changes at the two end faces of the the cavity (which are assumed to be close to zero), and *m* is the mode number of the Fabry-Pérot resonance. We used the dielectric function of alumina and template-stripped gold from experimentally measured data from references 1 and 2, respectively, and the dielectric function of silver was approximated by the Drude model.

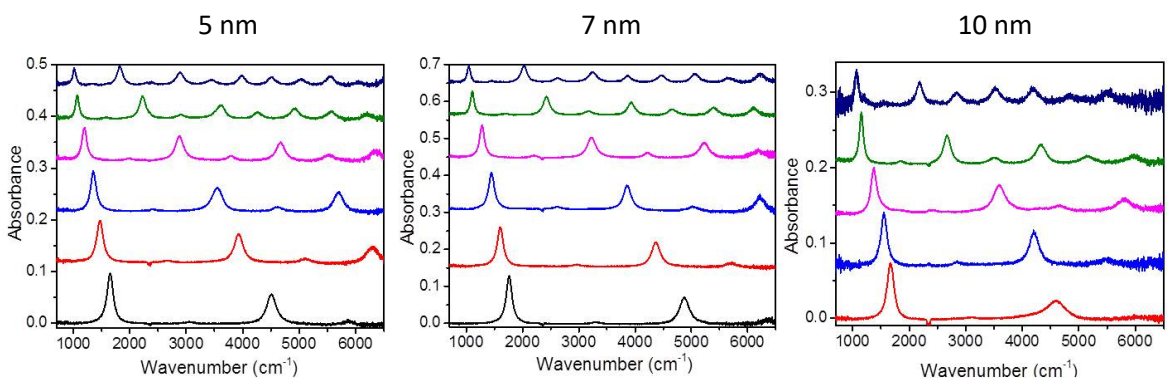

**Supplementary Figure S1.** FTIR spectra from 5 nm, 7 nm and 10 nm (from right to left) nanogap cavities in a 2D stripe pattern with different metal stripe widths: 500 nm (except for 10 nm nanogap cavity), 600 nm, 700 nm, 900 nm, 1300 nm, and 1700 nm (from bottom to top).

**Benzenethiol coating**

We used a solution of benzenethiol (BZT, Sigma-Aldrich), with 98% purity diluted to 2 mM in ethanol. Before putting the nanogap sample into the solution, the alumina inside the nanogap was partially removed by buffered oxide etchant (BOE). The BOE etching rate was around 1 nm per second, and we used an etching time that allowed the depth of etching into the gap to be close to nanogap width. The nanogap array sample and bare template stripped silver on a glass slide were soaked in the solution for 24 hours to allow BZT monolayer formation on the metal surface and sidewall inside the gap. Before measurement, the samples were cleaned with a flow of ethanol for 2 minutes to remove excess BZT molecules. The angle of the BZT molecules on the metal surface might be different depending on the crystalline directions of the metal. For simplicity, we assume the angle of the molecules is 90 degree, both on the flat surface and inside the gap. Due to the surface selection rule [17], we also assume that only the BZT molecules inside the gap will absorb light when the light normally incident on the sample surface with polarization perpendicular to (across) the gap. In case of the flat surface, oblique light is needed to provide the polarization of light that is the nearly perpendicular to the metal surface for vibrational absorption in the mid infrared.

**Surface-enhanced infrared absorption enhancement factor (EF)**

The SEIRA measurements from nanogap samples were compared with those obtained from reference samples (template-stripped flat silver surface) to calculate the IR absorption EF. We assume BZT molecules form a uniform monolayer on gold and silver surface in both nanogap and on top metal surface, with same packing density.

To take the reference signal from a bare silver surface, we used the main bench of a Nicolet Series II Magna-IR System 750 FTIR. Due to the surface selection rule, oblique infrared beam at incident angle 83° was used to maximize infrared absorption. The light spot area was an ellipse around 2 mm by 3 mm in conjugate and transverse diameter.

EFs were calculated for six absorption bands of BZT using the equation:

$$EF = \frac{A_{gap}}{N_{gap}} / \frac{A_{surf}}{N_{surf}}$$

where $N_{surf}$ is the number of BZT molecules contributing to the absorption on a bare template stripped silver film, $N_{gap}$ is the number of BZT molecules contributing to the SEIRA signal, and $A_{gap}$ and $A_{surf}$ are the intensities of the absorption band of interest in the gap and on the template stripped silver (area coverage ratio is one). Also, an angle correction $sin(83°)$ was added to the $A_{surf}$. So the equation becomes:

$$EF = \frac{A_{gap}}{D \times \frac{L}{P} \times 4wL} / \frac{A_{surf}}{sin(83°) \times D \times S_o}$$

where the $D$ is the surface density of BZT monolayer on a metal surface, which we assumed is the same for nanogap and template-stripped metal surface; $L$ (=130 um) is width and length of the square aperture in the microscope; $P$ (=3 $\mu$m) is the period of embedded nanogap; and $w$ is width of the nanogap. We assumed the depth of nanogap was the same as its width after BOE etching. So the surface area inside the nanogap for each embedded nanogap device is $4wL$. $S_o$ is the area of the light spot area in the FTIR.

| Absorption peak (1/cm) | Absorption on substrates and EF — Template stripped silver | Nanogap cavity width w=450 nm | EF | Nanogap cavity width w=600 nm | EF |
|---|---|---|---|---|---|
| 1575 | 3.00E-05 | 0.00238 | 5.49E+05 | | |
| 1473 | 3.30E-04 | 0.00752 | 1.58E+05 | | |
| 1181 | 1.20E-04 | | | 0.0109 | 6.28E+05 |
| 1073 | 1.20E-04 | | | 0.00168 | 9.69E+04 |
| 1022 | 1.00E-05 | | | 0.00234 | 1.62E+06 |
| 1000 | 1.00E-05 | | | 0.00172 | 1.19E+06 |

| Absorption peak (1/cm) | Nanogap cavity width w=700 nm | EF | Nanogap cavity width w=900 nm | EF | Nanogap cavity with graded width | EF |
|---|---|---|---|---|---|---|
| | | | | | | |
| | | | | | | |
| | | | | | 0.00077 | 4.46E+04 |
| 1073 | 0.00273 | 1.67E+05 | | | 0.00129 | 7.44E+04 |
| 1022 | 0.00278 | 1.94E+06 | 0.00242 | 1.66E+06 | 0.00226 | 1.56E+06 |
| 1000 | 0.00185 | 1.31E+06 | 0.00121 | 8.30E+05 | 0.00147 | 1.02E+06 |

**Supplementary Table S1.** SEIRA EF in different nanogap cavity devices.